# Development of a Medical Tele-Management System for Post-Discharge Patients of Chronic Diseases in Resource-Constrained Settings

Amusan, E. A.[1,*], Emuoyibofarhe, O.J.[1,2,3], Arulogun, O.T.[1]

[1]Department of Computer Science and Engineering, LadokeAkintola University of Technology, Ogbomoso, Nigeria.
[2]Centre for Excellence in Mobile e-Services, University of Zululand, South Africa
[3]Hasso Plattner Institute, University of Potsdam, Germany.
Corresponding author: [1,*]*eaadewusi@lautech.edu.ng*

**ABSTRACT**

Medical tele-management is an emerging field of study in telemedicine that proposes an interactive and proactive disease management approach which combines tele-monitoring and tele-consultation services through an information and communications technology (ICT) supported partnership. On one hand, chronic diseases require frequent and continuous monitoring to avoid complications; on the other hand is the need for a platform where health workers in rural settlements can consult or interact with their counterparts in urban areas to reduce isolation. However, telemedicine systems exist singly either as tele-monitoring or tele-consultation systems or majorly in developed countries with dedicated and adequate ICT resources and infrastructure. This work developed a combined tele-monitoring and tele-consultation system for the management of chronic diseases within an information and communication technology resource-constrained setting. This is achieved through the development of a multi-tiered framework and model deployed to provide medical tele-management services for rural African communities. The developed system achieved reliability score, availability, uptime and downtime of 0.9, 99.65%, 99.65% and 0.21% respectively. The evaluation results of the post-implementation task revealed that response means of 4.20 and 4.22 were achieved for the system degree of relevance (SDR) and system ease of use (SEU), respectively on a rating scale of 1 to 5.

## 1. INTRODUCTION

Chronic diseases such as congestive heart failure, diabetes, chronic obstructive pulmonary disease and stroke etc, is prevalent in developing and third world countries. Most developing countries in Africa are faced with limited medical personnel and inadequate quality health services to address the many health care problems. This is further worsened by the poor condition of the available health care facilities and the high cost of management of chronic diseases prevalent in Africa [1]. Many of these countries face a serious shortage of doctors, especially specialists, which results in the problem of lack of access to quality healthcare service by large percentage of the population and consequently, professional isolation of the few medical personnel available in these areas. Access to healthcare service can make a huge difference, that is, between life and death especially in emergency cases [2]. Several studies have shown that 31 African countries have fewer than 10 physicians per 100,000 people. Particularly in Nigeria, the physician per people ratio is averagely one to one thousand [3]. Telemedicine provides a tool for addressing and improving this situation.

Telemedicine, which literally is the delivery of healthcare services and exchange of health information over a distance, has been made known by several studies in the last two decades to have potential benefits and tremendous capabilities to bridge the gap of access to quality healthcare and reduce cost [1,4-5].Telemedicine systems contribute to continued care and patient education, assist patients in taking medications, and improve healthcare delivery [6].

Telemedicine covers a wide range of applications in areas such as Tele-consultation, Tele-monitoring, Tele-education, Tele-management (combination of Tele-monitoring and Tele-consultation), Tele-Oncology, Tele-dermatology, Tele-Ophthalmology and Tele-Radiology [7].

Medical tele-management is the combination of tele-monitoring and tele-consultation services for a more robust healthcare service package [8]. The synergy of both services offers the possibility to continuously monitor outpatients with chronic conditions for proper post-treatment management to avoid relapse and a deteriorating health condition and at the same time provide a platform where health workers/professionals in rural settlements can consult or interact with their counterparts in urban areas.

According to Rice [9], tele-monitoring is defined as the ability to measure, store and, as necessary, forward detailed information about the "vital signs" of a patient at any particular point in time. Continuous monitoring affords early detection of emergency situations that need quick intervention and it also aids in optimum allocation of healthcare services for those in need of it. Tele-monitoring is progressively being used in the management of chronic diseases as it facilitates timely transmission of patient' data [10].

Tele-consultation is an electronic communication between a physician and a client, another physician, or another health professional for the purpose of delivering health care services and information over small and large distances [11]. Transmission may include data, images, and/or voice. Most commonly, tele-consultation takes place between a health care professional (HCP) and a patient or between HCPs for diagnostic or therapeutic advice or for educational purposes. The goal of tele-consultation is to eliminate the barriers of distance and promote equal access to health services to remote areas where immediate consultations are unavailable.

Tele-monitoring has been reported to be an efficient and effective model in the management of chronic diseases [12]. When combined with tele-consultation, it offers a more robust solution for the management of chronic illnesses.

This paper presents a multifaceted medical tele-management system for the monitoring of multiple chronic disease conditions which is characterized by automatic transmission of patients' data (vital signs). Wireless body area network (WBAN) and web real time communication (webRTC) are some of the technologies used to provide services that reduced patient's journeys, hospital visits, and admissions, save the time of healthcare professionals, support individuals with chronic health conditions, and improve the quality or effectiveness of the health care or treatment that is delivered. The research was conducted using the ICT infrastructure at the LadokeAkintola University of Technology, Ogbomoso, Nigeria in conjuction with the University hospital.

## 2. RELATED WORKS AND TECHNOLOGIES

Gund [13] developed a disease management system which is Internet based for chronic heart failure. However, the system is only functional in monitoring patients and their vital signs within a home environment; mobility was not supported and does not enable Physician-Physician tele-consultation. Another limitation of this system is the fact that the patients being monitored would have to by themselves; periodically, manually enter their vital signs as input into the system.This could lead to transmission of wrong readings by patients and can be significantly misleading.

Mugoh and Kahonge [14] developed a telemedicine system for the management of blood Pressure (BP) among hypertensive patients. The beauty of this system is that it is mobile-phone based such that patients can transmit their BP condition to the doctors from the comfort of their residence using their mobile phones. The telemedicine system achieved an uptime of 98.46%, downtime of 0.37%, reliability score of 0.9 and availability of 99.82%. However, data collection in this mobile telemedicine system was done manually; patients would have to enter their vital signs (BP readings in this case) to the interface of the mobile application at intervals of the day. Further work suggested automatic transmission of the BP readings from the Wrist BP Monitors to the android application on the mobile phone via Bluetooth technology.

Amusan, Emuoyibofarhe and Arulogun [15] designed a conceptual framework for a medical tele-management system for ICT resource-constrained settings. They presented the design of multi-layered service oriented architecture for efficient and effective management of chronic diseases but with no implementation, which was proposed as a future work.

Setyono, Alam and Soosai [16] presented a tele-consultation system that allows the citizens and doctors of Malaysia to interact and exchange information related to dengue fever. The developed system was expected to assist and ease the dengue monitoring and predicting the dengue outbreak cases in Malaysia.

Zhang, Thurow and Stoll [17] developed a knowledge-based tele-monitoring system for remote health care to address the problem of scalability using context-aware middleware. This was to ensure that a new sensor device can be easily added to the system with relative ease.

Gennaro, Donadeo, Bulzis, Ricci, Citarelli, Resta, Natale, Brunetti and Ospedale [18] developed a tele-monitoring and tele-consultation system for the management of patients with chronic heart failure. It presents an integrated model of telemedicine for the follow up of patients with such conditions. However, the system is limited to manage heart failure alone; other chronic illnesses may not be monitored or managed.

In conclusion, arising from the related works reviewed in this sub-section, it is evident that not so much has been done to combine both services (that is, tele-monitoring and tele-consultation) to give a robust medical tele-management system.

Therefore, this work developed asystem which combinestele-monitoring and tele-consultation for the management of chronic diseases. It addressed the limitations in scope of implementation of existing systems by providing the capability to facilitate both patient-to-physician and physician-to-physician interaction and at the same time provide multimedia (audio, image and video) communication. Furthermore, it supports automatic transmission of patients' vital signs at a reduced downtime and improved uptime. An added advantage is that it is also a multi-disease management system as patients can have a combination of chronic diseases at a particular point in time. This will significantly reduce the burden and time needed in face to face consultation and invariably reduce the cost burden of managing chronic diseases.

## 3. METHODOLOGY

This research adopted a design and experimental approach for the development of the medical tele-management system.

### 3.1 Requirements definition of the developed system

The design requirements of the developed tele-management system are divided into domain requirements - healthcare; tele-monitoring and tele-consultation, application requirements - functional and non-functional requirements and deployment environment requirements - network and operational requirements.

**a) Domain requirements**

Since the system is domain-centred, it could be either tele-monitoring or tele-consultation. Top requirements in both domains are availability, reliability and data security (confidentiality, integrity and authenticity).

**b) Application Requirements**

These are described in terms of both functional and non-functional requirements. The following requirements are necessary: (i) connection to the biomedical sensor (ii) Read patient's vital signs (iii) upload vital signs (iv) analysis of data streams (v) the trigger of emergency alerts (vi) initiation of a tele-consultation session (vii) termination of a tele-consultation session (viii) access to EMR.

**c) Deployment environment requirements**

This puts into consideration the requirements the system needs to be able to work successfully in the location of deployment. It is particularly aimed at meeting the deployment environment constraints which the existing frameworks could not satisfy at all or as desired.

This was done through the use of devices that are less expensive and rechargeable making it advantageous for rural or resource-constrained areas which are characterized by shortage of finance, inadequate power supply to access healthcare services. These devices are handy, mobile and easy to operate, thus eliminating the need for sophisticatedapplications that require highly skilled manpower which is limited in availability in these areas.

## 3.2 Overall System Architecture

With a special consideration for user roles and activities, as well as scenarios of possible use and users' expectations for a system like this, patients and medical practitioners are the intending major users of the system and they can be in any of the possible locations as shown in the system architecture represented by Figure 1.

A patient can access the system in the comfort of his home, office, market or even on a public transport. This is made possible by the integration of mobility in the design of the system as humans are nomadic in nature. At the core of the framework lies the data center situated in a clinic or just a dedicated data center in a community and accessible by a group of users in that community. In this case, an assisting operator can render help, especially where there needs to be a little more sophistication in the tele-consultation session. These centers then have to be supplied with relatively simple clinical measurement devices necessary to obtain the basic medical data like blood pressure, temperature, pulse rate, weight, etc to support tele-consultation. The role of the staff is to capture data from the patients, send them to the consulting specialist in real-time via the text chat or audio-visual devices. It is important that the data centers or telemedicine centers be so strategically located and be accessible by users. The consulting medical practitioner also has access to the system via any of the two previously discussed deployment modes.

The complete functional structure of the framework is summarized as follows: an eligible system user who can either be a patient or a medical staff will login into the system after registration. The patient uses a Patient ID to log in while the staff uses the staff ID to log in, all of these are communicated to the respective users at the point of registration. If the login is successful, then the system acquires the required permission which confirms whether the person attempting to login in is a staff or a patient. If the user is a staff, the system loads the appropriate modules for the staff like Add/Update Staff Profile and Add/Update Patient Profile Module. If the user is a patient, he/she is presented with an interface to select service type which can either be Tele-monitoring or Tele-consultation.

The framework has been so designed to allow remotely located patients to be monitored. It also allows them to initiate consultation with a physician in another geographical location. The various devices such as the mobile phone, PDA, laptop or even a desktop computer are clients needing a service. Authentication into the system is via patient identification number (patient ID) generated at the point of registration. Authorization is by Role-Based Access Control (RBAC) to specify access rights and privileges to the system. Security considerations were based on the open Secure Socket Layer (SSL) library and Public Key Infrastructure (PKI) management.

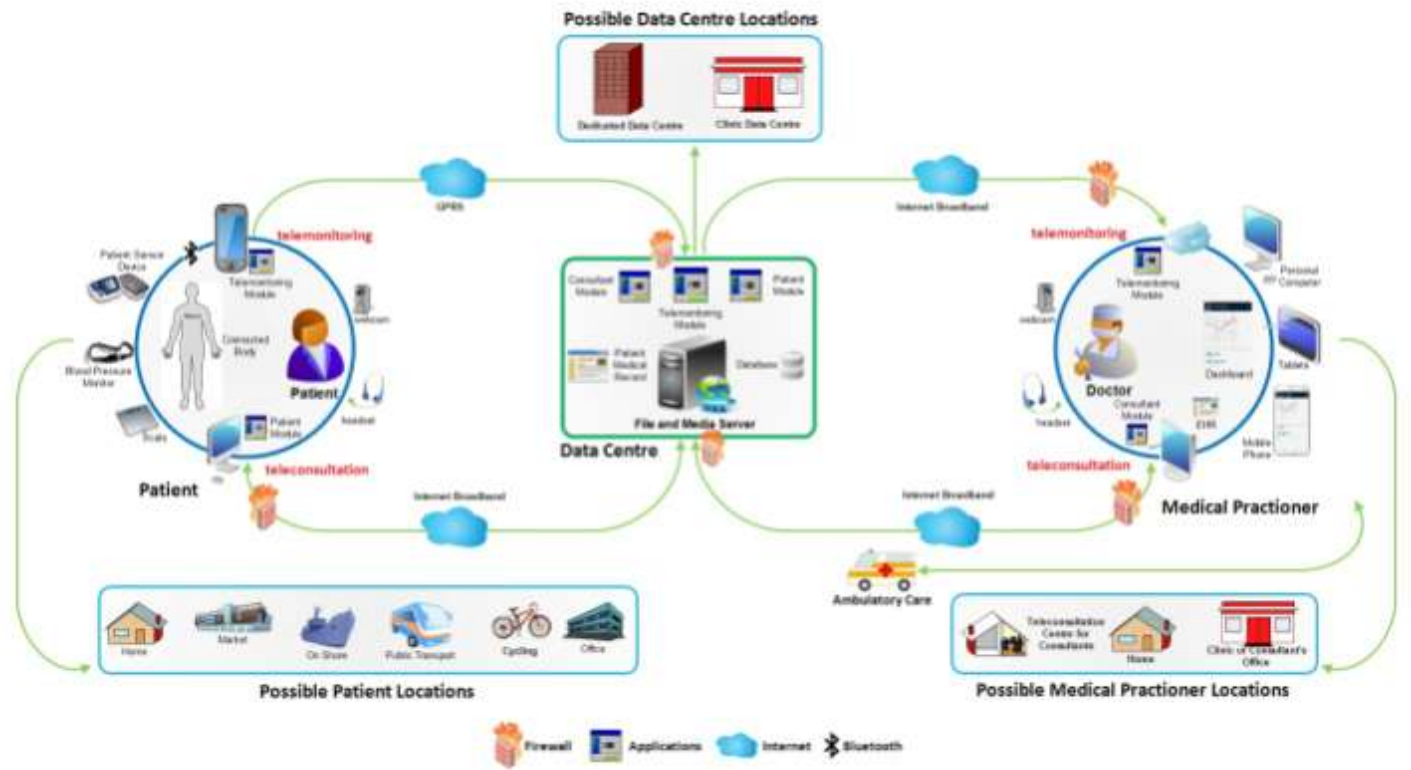

Figure 1: The Tele-management system architecture

The tele-monitoring application was modeled around the 3-tier client-server architecture developed using body area network. The first tier is made up of the client comprising biomedical sensors attached to the body of a patient whose health condition requires continuous monitoring. The choice of biomedical sensors used determines the physiological signs to be measured. This research used the fit bit medical biosensing device as shown in Figure 3(b) as it is a multi-parametric sensor. However, the system was designed such that any other wearable sensing device can interface with it to monitor other chronic conditions given that the device is Bluetooth-enabled. This fit bit then relays the medical information (i.e, heart rate and blood pressure ) to the second tier of the sub-system using the IEEE 802.15.1 (Bluetooth) wireless communication protocol. Tier 2 is the smartphone (android-based in this context) which serves as the data hub as it provides more computational capability for storage and processing before being sent to the web-based platform via the Internet using GSM/GPRS. The third tier is the medical network which is the back end that receives the information. Figure 2 shows the communication that takes place among these tiers within the tele-monitoring module.

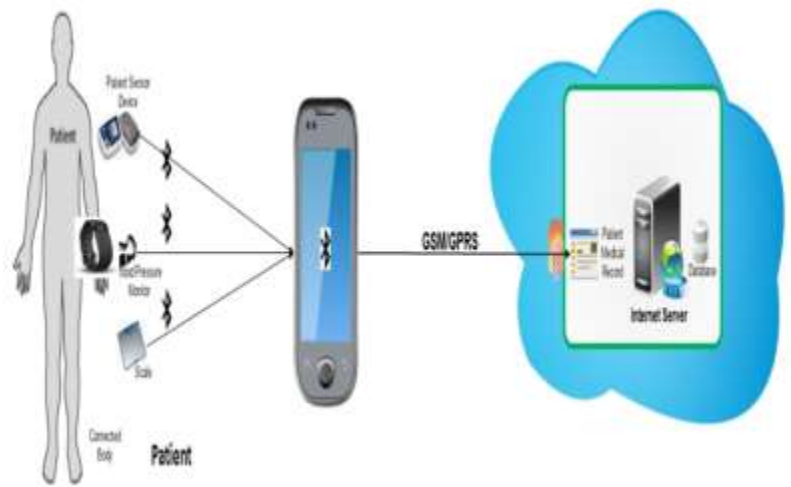

Figure 2: Communication Channel for the Tele-monitoring Sub-system

Medical tele-monitoring entails continuous monitoring and transmission of vital signs like heart rate and blood pressure. After due consultations with some doctors in the university clinic, threshold values for the tele-monitored parameters were established, which are subject to modification by the health care expert according to the specific peculiarities of the patient. These thresholds are necessary for the detection of abnormality in condition. Table 1 shows the established thresholds for the tele-monitored parameters, that is, heart rate, systolic and diastolic blood pressure. The moment the monitored vital data falls outside the established limits, alerts will be triggered both on the mobile phone and the medical server.

Table 1: Established Thresholds for the Tele-monitoring Parameters

| Tele-monitored Parameters | Established Threshold Values |
|---|---|
| Systolic blood pressure | <50 or >100 beats per minute (bpm) |
| Heart rate | <100 or >160 millimeters of mercury (mmHg) |
| Diastolic blood pressure | <60 or >95 mmHg |

### 3.3 Devices Used for Achieving Tele-management

For the tele-monitoring aspect of this work, two (2) biomedical sensing devices were acquired and used, they are: Blood Pressure Monitor by A&D and Fit bit charge HR device. The blood pressure monitor is used to measure the blood pressure of the patient. Very high or low blood pressures suggest imminent danger upon the patient under tele-monitoring and it is measured in millimeters of mercury (mmHg). The Fit bit device senses patients' vital signs like the heart rate measured in beats per minute (bpm). It is also capable of measuring the activity level and sleep rate of the monitored patients. The readings were communicated at periodic intervals to the mobile phone via the Bluetooth protocol and further transmitted to the medical server; the devices are as shown in Figures 3a and b respectively.

The tele-consultation aspect of this work was experimented and achieved in two modes. The first mode was through the use of the inbuilt webcam and internal microphone of the personal computer. The second mode was the use of external device like the two-way audio surveillance camera by TP-LINK whose diagram is as shown in Figure 3c. This second mode is necessary to cater for desktop systems and monitors without webcams. The 2-Way Audio Surveillance Camera (TL-SC3130) is such that it allows two clients to connect, rendering both audio and visual functionalities which are used to facilitate tele-consultation. The client on one end connects to the user camera, and the user also connects to the client's camera, each using the other's IP address to connect.

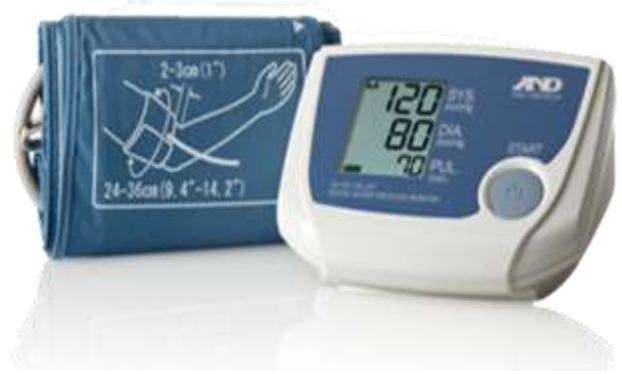

(a): The acquired blood pressure monitor (UA-767-PBTCi)

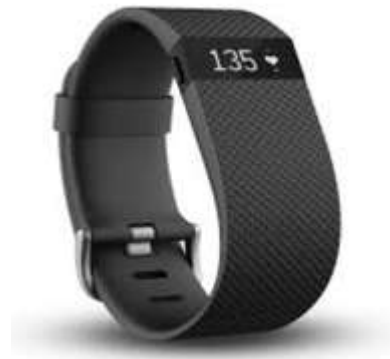

(b) Fit bit ChargeHR (FB405)

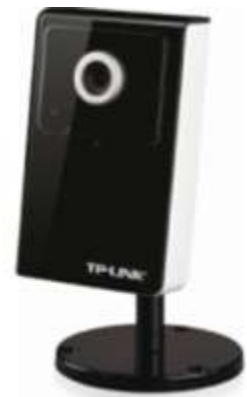

(c) 2-Way audio surveillance camera (TL-SC3130)

Figure 3(a) – (c) : Acquired devices: (a) blood pressure monitor, (b) Fit bit Charge (c) 2-Way audio surveillance camera

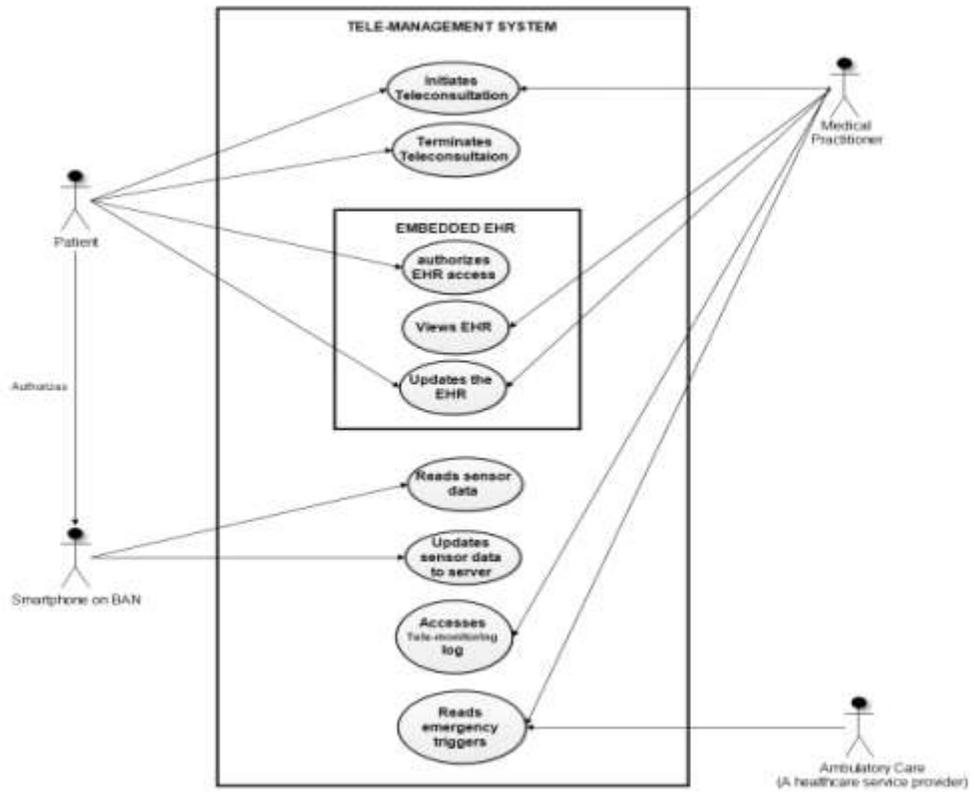

Figure 4: Use case diagram for the system

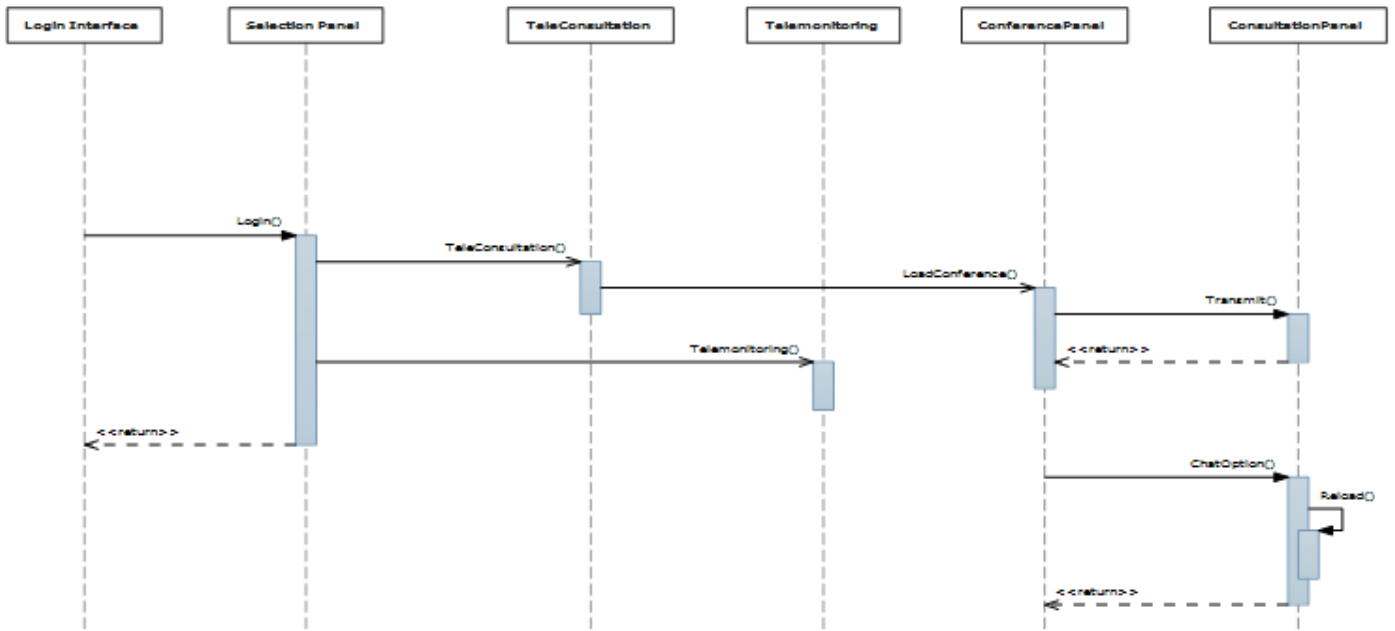

Figure 5: Sequence diagram for the system

## 3.4 System Analysis Modeling with Unified Modeling Language (UML)

A model of the designed system was built using Unified Modeling Language (UML). In this paper, UML Use case diagram and sequence diagram were used to model the system as shown in Figures 4 and 5 respectively.

The use case diagram comprises two principal actors: medical practitioner and Patient. The Practitioner has the privilege to initiate and terminate tele-consultation sessions,, view and update patients' record as well as respond to emergency triggers. The patient can initiate and terminate sessions, authorize access to his/her record, read, as well as update sensor data. The sequence diagramis as shown in Figure 5. A user is required to login at the login interface and uses the selection panel to select a service, either tele-monitoring or tele-consultation service.

## 4.0 SYSTEM IMPLEMENTATION

### 4.1 The Developed Medical Tele-management System

The developed system is composed of two (2) basic applications, namely: tele-monitoring and tele-consultation applications to validate the developed architecture. This was achieved by setting up an experimental test-bed for live streaming from which live data were captured. The tele-monitoring application was modeled around the 3-tier client-server architecture developed using wireless body area network (WBAN). The tele-consultation segment of the implemented tele-management system supports image, voice and video based consultation and this is made possible by web real time communication (web RTC).

The system is multi-access based and so, client applications can access the server application either via a windows mini browser on a smartphone or a laptop/desktop.

The entire system was implemented as web services. The interfaces of the various client applications developed connect to the web server via some deployed web services. These services ensure that a system user connects seamlessly to the server without being exposed to the complexities of the process. Figure 6 shows the web service interface for the system. The methods exposed here are such that enable the patient and healthcare service provider to get different services from the tele-management application server.

Figures 7(a) and (b) are the home page and login screen respectively. The home page is the first port of call to any user of the system. However, the only noticeable link here is the Login button. This is so because the system adopts the role-based access mechanism to assign responsibilities to the users who can either be a patient, medical expert or administrator. The moment the login link is clicked, the system gives access based on the user's job function. This measure is put in place to secure the system both internally and externally.

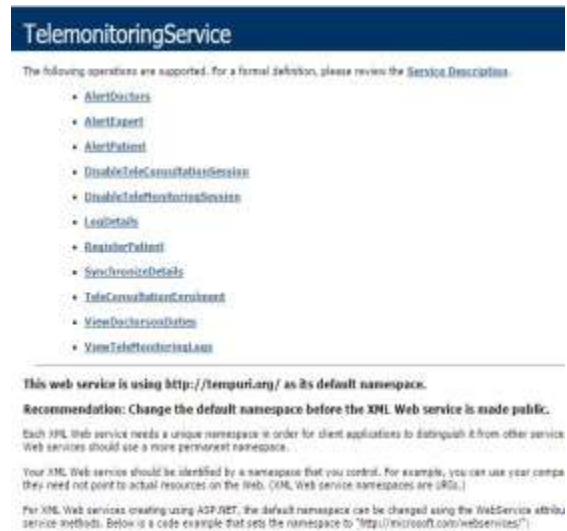

Figure 6: System Web service interface

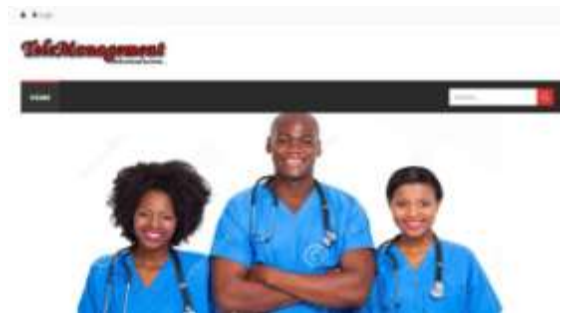

Figure 7(a): System home page

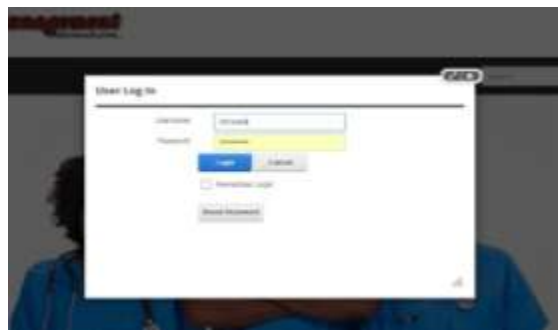

Figure 7(b): System log in interface

Some screen shots of the mobile application are depicted in Figures 8-10. Where Figure 8 represents the log-in interface of the tele-monitoring sub-system, Figure 9 captures the screen of the android-based phone that is connected to the Fitbit biomedical sensor. Figure 10 shows a short message service (SMS) received by the Physician who is remotely monitoring a patient.

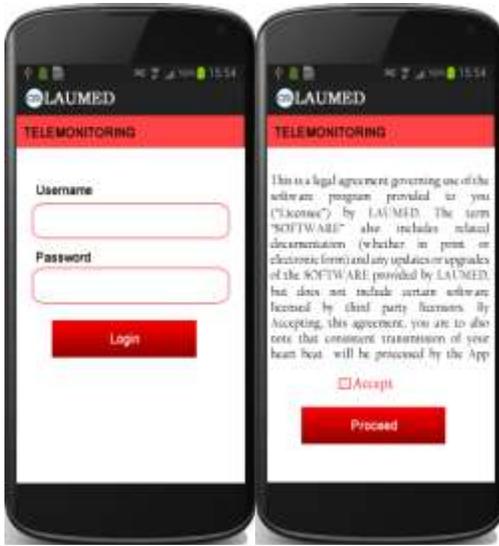
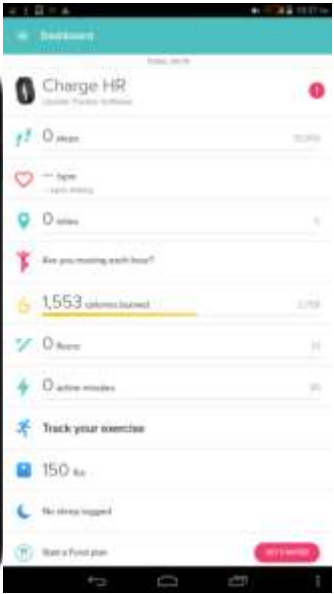

Fig. 8: Login interface of the tele-monitoring subsystem (Patients' side)

Fig. 9: Smart phone connected toFitbit device

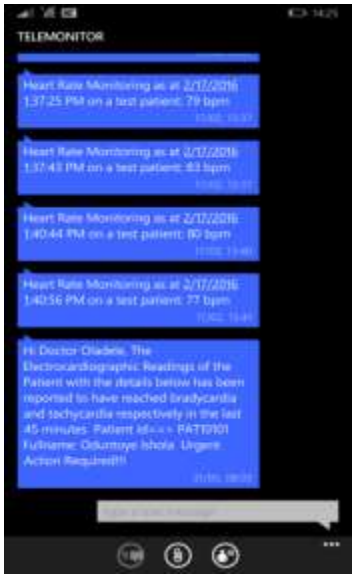

Figure 10: Physician receiving an SMS alert

The tele-consultation sub-system implemented in this work provides an interactive two-way remote consultation among all system users, that is, patients in the rural areas, distant rural health care givers and specialists in the urban cities by using chatting approach with audio-visual capabilities over the Internet. This was done to widen and extend the scope of existing systems. The implementation of the tele-consultation application is in two modes. The first is Patient-Physician interaction while the second mode is the Physician-Physician interaction.

**Patient-to-physician tele-consultation**
For patient-to-physician communication, interaction can be audio-visual. Figure 11 is the first interface presented to the patient, while the chat options could also be selected from there. Figure 12 is the next interface supposing the patient chooses to interact by chat.

**Physician-to-physician tele-consultation**
A part of the objectives of this work was to develop a system that can reduce or totally eliminate the geographical isolation among physicians in resource-starved location giving them the opportunity to consult and seek second opinion from those in resource-rich settings. This section satisfied that objective; for the physician-to-physician communication, two types of tele-consultation are supported by the system. They aretele-consultation by image and tele-consultation by video.

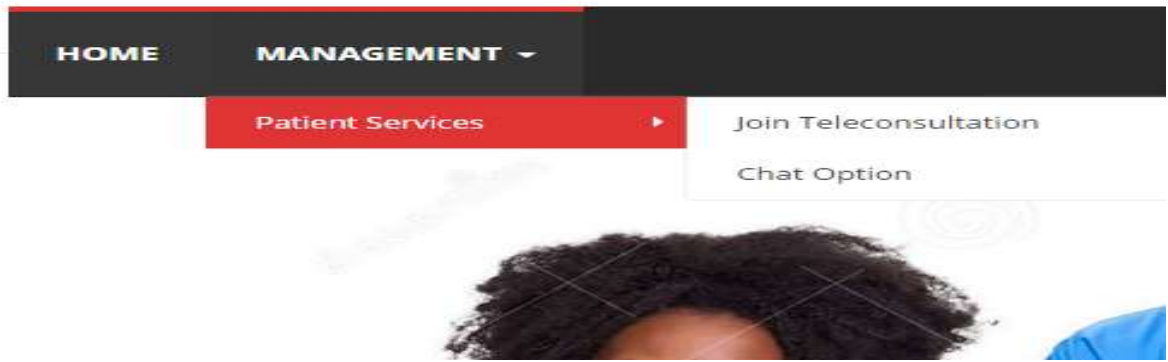

Figure 11: Patient launching tele-consultation application

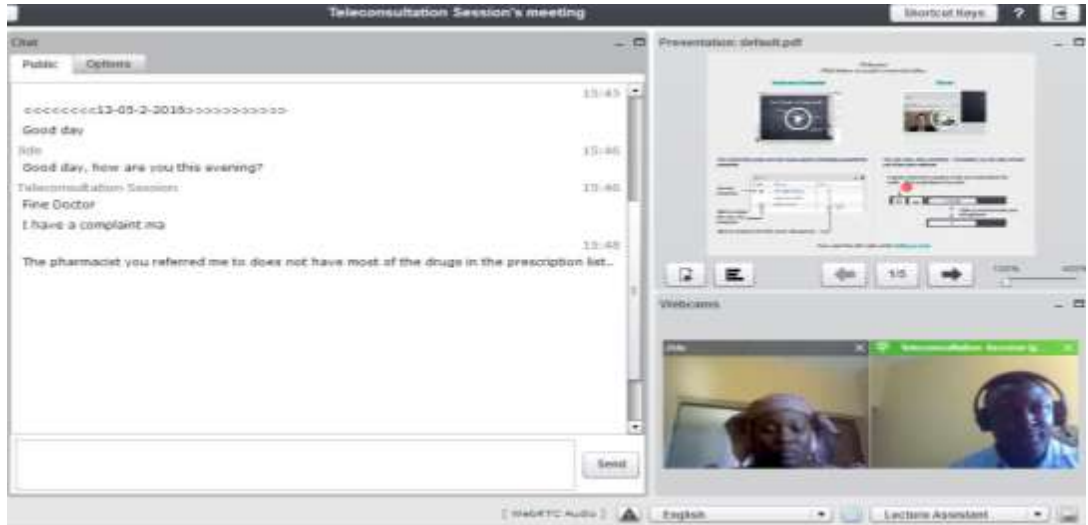

Figure 12: Patient-to-physician tele-consultation interface supporting audio-visual communication via chat

### 4.2 System Evaluation

Performance evaluation of the developed system was done in two phases. First is the objective (system-centric) evaluation which was done by evaluating the system for reliability, availability, uptime and downtime metrics. The second phase is the subjective (user-centric) evaluation. System Ease of Use (SEU) and System Degree of Relevance (SDR) metrics were evaluated based on users' assessment using a tailor-made Likert rating scale questionnaire.

For the objective evaluation, the developed system was evaluated for reliability and availability. Table 2 is the result of the system performance analysis based on reliability. The following metrics were used to measure the reliability of the system: uptime, downtime and mean time before failure (MTBF). To compute for reliability,

$$R = 1 - \lambda_T \qquad 1$$

where $\lambda_T$ is known as the failure rate. Failure rate is measured as the number of failure in a specified period of time. During the test period which lasted for a total of forty-eight (48) hours, the total number of failures encountered is six (6) from Table 2. Substituting these values into the equation 1, the reliability of the system is as shown below:

$$R = 1 - \frac{6}{48} = 1 - 0.1 = 0.9$$

From computation, the system achieved a reliability score of 0.9.

Similarly, the following metrics were used to evaluate the availability of the system: expected working time, uptime, response failure, mean time before failure (MTBF) and mean time to recover (MTTR).

To compute the system availability $A_T$,

$$A_T = \frac{Uptime}{Uptime + Downtime} \times 100\%$$
$$= \frac{2870}{2870 + 10} \times 100\%$$
$$= \frac{2870}{2880} \times 100\% = 99.65\%$$

The overall system availability is 99.65%.

Table 2: System reliability evaluation analysis

| | Test Period (hrs) | Uptime (mins) | Downtime (mins) | No of failures | MTTR (mins) | MTBF (mins) |
|---|---|---|---|---|---|---|
| **Day 1** | 8 | 479 | 1 | 1 | 1 | 462 |
| **Day 2** | 8 | 478 | 2 | 1 | 2 | 465 |
| **Day 3** | 8 | 476 | 4 | 2 | 4 | 235 |
| **Day 4** | 8 | 478 | 2 | 1 | 2 | 471 |
| **Day 5** | 8 | 479 | 1 | 1 | 1 | 475 |
| **Day 6** | 8 | 480 | 0 | 0 | 0 | 480 |
| **Overall** | 48 | 2870 | 10 | 6 | 10 | 2588 |

The system achieved a reliability of 0.9 score. This implies that the probability that the entire system will function as required without failure(s) within a period of ten (10) hours is 0.9. This means that out of a total of 10 hours of operation, a failure is only likely to occur in the 9th hour. However, this likelihood is independent of the system; it is rather a network failure.

Similarly, the system achieved an availability of 99.65 (two-nines). This implies an acceptable downtime of 0.35 that translates into 3.65 days per year as it does not affect the usability of the system. Also, after computation, the system achieved an uptime of 99.65%.

These evaluation results show that the system compares favorably well with that of Mugoh and Kahonge [13] who achieved uptime, downtime, availability and reliability values of 99.46%, 0.37, 99.82% and 0.9, respectively.

It is also observed from Table 2 that, day six (6) yielded a perfect result with no downtime, maximum uptime and zero failure rates. This is so because day six was actually a weekend with few staff and students on the university campus, the deployment environment, resulting in little contention for network resources.

The developed system was also subjectively evaluated using the questionnaire-based usability evaluation technique. A five-levelLikert scale was used – strongly disagree, disagree, neutral, agree and strongly agree. The Likert items (each question asked in the questionnaire) represent the usability dimension of the developed tele-management system.

A total of fifty (50) responses were received from potential users and data from the filled copies of questionnaire were captured, compiled and analyzed using Microsoft Excel 2010. The outcome of the data analysis of the Likert items is presented in the appendix.

The response mean in percentage was used to rate users' satisfaction with most of the features of the developed system. Both values are in the upper classes of the rating scale.

A summary of both evaluation metrics is presented in Table 3. The ratings revealed that the System Degree of Relevance (SDR) and System Ease of Use (SEU) are 88% and 90.67% respectively. Similarly, the system achieved response means of 4.20 and 4.22 for the SDR and SEU respectively on a rating scale of 1 to 5. It has a high and appreciable degree of relevance in reducing or eliminating the cost and stress associated with paying visits to the hospital for health condition monitoring. The developed system is also relevant in solving the problem of geographic isolation among physicians in remote locations. Furthermore, users find the system relatively easy to learn and use as they were able to complete their task in reasonable amount of time.

Table 3: Summarized users' assessment metrics for the developed system

| **Evaluation Criteria** | **Response in %** | **Response Mean** |
|---|---|---|
| System Degree of Relevance (SDR) | 88 | 4.20 |
| System Ease of Use (SEU) | 90.67 | 4.22 |

# 5. CONCLUSION AND FUTURE WORK

The hybrid telemedicine system developed in this paper has sought to make monitoring of patients with chronic diseases easier as it combined both tele-monitoring and tele-consultation services for a more robust healthcare package. It also supportsautomatic transmission of patient's vital signs to the physician for more accurate and efficient monitoring.The system provides an interactive two-way remote consultation among all system users, that is, patients in the rural areas, distant rural health care givers and specialists in the urban cities by using chatting approach with audio-visual capabilities over the Internet. This was done to widen and extend the scope of existing systems.Future work in the direction of this research can be pitched towards the development of data mining techniques for the data electronically generated and stored.

# APPENDIX

## DATA ANALYSIS OF THE ADMINISTERED RESEARCH QUESTIONNAIRE

| | Likert Items | Strongly Disagree | Disagree | Neutral | Agree | Strongly Agree | Response Mean | Response Mode |
|---|---|---|---|---|---|---|---|---|
| Q3 | You were able to complete your task with the system in a reasonable amount of time. | 0 | 2 | 2 | 20 | 26 | 4.40 | 5 |
| | | 0% | 4% | 4% | 40% | 52% | | |
| Q4 | Would you agree that the system is easy to learn? | 0 | 1 | 3 | 30 | 16 | 4.22 | 4 |
| | | 0% | 2% | 6% | 60% | 32% | | |
| Q5 | Would you agree that this system could reduce/eliminate problems of cost and stress of travelling associated with physical visit to a hospital? | 1 | 1 | 3 | 25 | 20 | 4.24 | 4 |
| | | 2% | 2% | 6% | 50% | 40% | | |
| Q6 | Would you agree that a full implantation of this system will reduce/eliminate frequent hospital re-admissions? | 0 | 1 | 8 | 23 | 18 | 4.16 | 4 |
| | | 0% | 2% | 16% | 46% | 36% | | |
| Q7 | Would you agree that this system will solve the problem of lack of access to healthcare services by patients irrespective of their location? | 1 | 2 | 1 | 29 | 17 | 4.18 | 4 |
| | | 2% | 4% | 2% | 58% | 34% | | |
| Q8 | Would you agree that this system will solve the problem of geographic isolation among physicians in remote locations? | 1 | 1 | 2 | 30 | 16 | 4.30 | 4 |
| | | 2% | 2% | 4% | 60% | 32% | | |
| Q9 | Would you support the idea of patients' involvement in the management of their own health through the use of mobile phones, computers and the Internet? | 1 | 2 | 7 | 25 | 15 | 4.02 | 4 |
| | | 2% | 4% | 14% | 50% | 30% | | |
| Q10 | Would you agree that this system is easy to use? | 2 | 2 | 2 | 30 | 14 | 4.04 | 4 |
| | | 4% | 4% | 4% | 60% | 28% | | |
| Q11 | Would you agree that this system, if developed on a national scale, could reduce mortality rate in Nigeria? | 0 | 2 | 2 | 26 | 20 | 4.28 | 4 |
| | | 0% | 4% | 4% | 52% | 40% | | |